\documentclass[aps,prd,superscriptaddress,nofootinbib,showpacs]{revtex4}

\usepackage{graphicx}

\usepackage{epsfig}
\usepackage{color}

\usepackage{amsmath}

\usepackage{amssymb}

\begin{document}

\title{Naked Singularity Formation In Brans-Dicke Theory}

\author{Amir Hadi Ziaie}
\email{am.ziaie@mail.sbu.ac.ir} \affiliation{Department of Physics, Shahid
Beheshti University, Evin, Tehran 19839, Iran}

\author{Khedmat Atazadeh}
\email{k-atazadeh@sbu.ac.ir} \affiliation{Department of Physics, Shahid Beheshti
University, Evin, Tehran 19839, Iran}

\author{Yaser Tavakoli}
\email{tavakoli@ubi.pt} \affiliation{Departamento de F\'{\i}sica, Universidade da
Beira Interior, Rua Marqu\^{e}s d'Avila e Bolama, 6200 Covilh\~{a}, Portugal}

\date{\today}

\begin{abstract}
Gravitational collapse of the Brans-Dicke scalar field with
non-zero potential in the presence of matter fluid obeying the
barotropic equation of state, $p=w\rho$ is studied. Utilizing the
concept of expansion parameter, it is seen that the cosmic
censorship conjecture may be violated for $w=-\frac{1}{3}$ and $
w=-\frac{2}{3}$ which correspond to cosmic string and domain wall,
respectively. We show that physically, it is the rate of collapse
and the presence of Brans-Dicke scalar field that govern the
formation of a black hole or a naked singularity as the final fate
of dynamical evolution and only for these two cases the
singularity can be naked as the collapse end state. Also the weak
energy condition is satisfied by the collapsing configuration.

\end{abstract}

\maketitle

\section{Introduction}

The process of gravitational collapse has been studied since about
60 years ago as a solution of Einstein field equations. When a
sufficiently massive star many times the size of the Sun exhausts
its nuclear fuel without reaching an equilibrium state such as a
neutron star or white dwarf, it collapses under its own pull of
gravity at the end of its life cycle. Therefore, gravity overtakes
and dominates the other three forces of nature, in particular, the
weak and strong nuclear forces, which generically provide the
outward pressure in a star to balance it against the inward pull
of gravity. In such ultra-strong gravity regions the densities and
spacetime curvature diverge and a spacetime singularity will be
born, either hidden within an event horizon (a black hole) or
visible to the external universe (a naked singularity), as
predicted by the singulary theorems in general relativity. The
visibility or otherwise of the singularity to outside observers is
determined by the causal structure of the dynamically developing
collapsing cloud as governed by the Einstein's field equations.
When the internal dynamics of the collapse delays the formation of
the horizon, the singularity becomes visible, and may communicate
physical effects to the external universe \cite{01}.

Up until now a great deal of effort has gone into the study of the
nature of singularities and large classes of the solutions of
Einstein's field equations treating singularities have been
represented. The general and exact class of solutions of
Einstein's field equations describing spherically symmetric
pressureless matter (dust) for motion with no particle layers
intersecting, independent of the homogeneity assumption, was
originally introduced by Lema$\hat{i}$tre \cite{02, 03} in an
attempt to describe cosmology, which was further developed and
studied by Tolman \cite{04, 05} and Bondi \cite{04,06}. This class
could be used to model the gravitational collapse of matter from
general inhomogeneous initial conditions and the end state of
gravitational collapse of a massive star can be studied within
this framework \cite{04}. Contrary to the collapsing Friedmann
case, in which the physical singularity occurs at a constant epoch
of time, namely at $t=0$, the singular epoch (the time $t =
t_{0}(r)$ at which the area radius of the collapsing shell of
matter at a constant value of the co-moving coordinate r vanishes)
in LTB model is a function of r as a result of inhomogeneity in
the matter distribution. This model can exhibit two kinds of naked
singularity: a shell-crossing \cite{07, 08} and a shell-focussing
singularity \cite{07, 09, 10}. These kinds of naked singularity
are generic for spherical dust collapse and with proper choice of
initial data they are globally naked. The main reason which causes
this model to be deficient is that it neglects the pressure, which
is likely to diverge at these singularities (where the density
diverges) \cite{07}. A special case of these classes of solutions
which has served as the basic paradigm in black hole physics is
the Oppenheimer-Synder study of a completely homogeneous,
pressure-free and spherically symmetric dust cloud collapse, where
a dust cloud undergoes a continued collapse which commences from
regular initial data (i.e., there is no trapped surfaces forming
at the initial spacelike surface from which the collapse begins
and the light rays can escape from the surface of the star to
faraway observers) to form a black hole \cite{04, 11, 12}. Most of
our knowledge about the gravitational collapse is still based on
that model. Also it describes well the formation of the horizon
and evolution of the central, space-like singularity. Oppenheimer
and Snyder analyzed the causal structure of the solution by
considering in particular an observer on the surface of the dust
cloud sending signals to a faraway stationary observer at
regularly spaced intervals as measured by his own clock. They
discovered that as the radius of the dust cloud approaches $2M$,
the spacing between the arrival times of these signals to the
faraway observer becomes progressively longer, tending to
infinity. This effect has since been called the infinite redshift
effect. The observer on the surface of the dust cloud may keep
sending signals after its radius has become less than $2M$, but
these signals can not escape from this region since the speed of
light is the limiting propagation velocity for physical signals.
within a finite affine parameter interval, these signals reach a
true singularity at $r = 0$ \cite{13}. This singularity is formed
inside a black hole, as a final state of the collapse process and
also accords to the concept of the cosmic censorship. The cosmic
censorship conjecture which is initially represented by Penrose
 \cite{14}, explain the properties of the final singularity of a
gravitational collapse. The main result of this conjecture is that, all spacetime
singularities arising from regular initial data that appear in a gravitational collapse
(in an asymptotically flat universe) are always surrounded by an event horizon
and hence invisible to outside observers (no naked singularities). Moreover in the
strong version of this conjecture, such singularities are not even locally naked,
i.e. there is not any timelike or null geodesics which can emerge from these
singularities and go to future infinity \cite{07}. This hypothesis plays a
fundamental role in both the theory and applications of black hole physics and
has been recognized as one of the most important open problems in classical
general theory of relativity. Up to now many exact solutions of Einstein's field
equations with several kinds of field-sources which admit naked singularities
depending upon the nature of the initial data and the kinematical properties near
the singularity have been considered. The models studied so far include the
collapse of scalar fields \cite{15, 16} as well as other matter sources including
dust \cite{17, 18, 09}, radiation \cite{17, 19, 10}, perfect fluid \cite{17, 20, 07},
imperfect fluids \cite{17, 21}, and null strange quark fluids \cite{15, 22}.

One would like to inquire about what are the possible physical factors operating
during the process of a continual collapse of a massive matter cloud which
results in the formation of a naked singularity or a black hole as the collapse end
state. Such an investigation should help us understand much better the physics
of black holes and naked singularity formation in gravitational collapse. It was
recently shown \cite{23, 24} that for the spherical dust collapse the shearing
effects and inhomogeneity present within the collapsing cloud do play a crucial
role in delaying the formation of the trapped surfaces and the apparent horizon.
These effects could in fact make the geometry of apparent and event horizons
distorted sufficiently, which exposes the singularity to the external observers
\cite{23}.

General relativity is not the only gravitational theory which
explain the gravitational phenomena. There are alternative
theories of gravity that can explain the gravitational phenomena
and lead to many interesting results and physical interpretations.
One of those, is the scalar-tensor theory of gravity, which has
been actively studied as an alternative and successful theory of
gravity. Recently, they have attracted much attention, in part
because they emerge naturally as the low-energy limit of many
theories of quantum gravity, such as Kaluza-Klein
theories\cite{25, 26} and supersymmetric string theories\cite{25,
27}. These theories are also important for "extended" cosmological
inflation models\cite{25, 28}, in which the scalar field allows
the inflationary epoch to end via bubble nucleation without the
need for fine-tuning cosmological parameters(the "graceful exit"
problem)\cite{29}. In Brans-Dicke theory, the simplest of
scalar-tensor theories, gravitation is described by a metric
$g_{\mu\nu}$ and a scalar field $\phi$ coupled to both matter and
space time geometry which obeys a wave equation with a source term
determined by the matter distribution. Gravitational collapse in
this theory has been studied in some of the recent literature. In
this paper we study the formation of naked singularity as the end
state of gravitational collapse of a matter fluid obeying the
barotropic equation of state, $p=w\rho$ in the context of
Brans-Dicke (BD) theory. Being motivated by this concept, we
investigate the violation of cosmic censorship conjecture in the
collapse procedure of such a matter fluid.

\section{Basic equations}
In the context of Brans-Dicke theory with a self interacting potential and a matter
field, the action is given by
\begin{equation}\label{eq1}
{\mathcal S}=\frac{1}{2\kappa}\int d^{4}x\sqrt{-g}\left(\Phi
R-\frac{\omega}{\Phi}\nabla^{\alpha}\Phi\nabla_{\alpha}\Phi-V(\Phi)\right)+
{\mathcal S}_{m},
\end{equation}
where the constant $\omega$ is the Brans-Dicke parameter and
$\Phi$ is the Brans-Dicke scalar field, which is related to the
gravitational "constant" by $G=\frac{1}{\Phi}$. Extremizing the
action yields the field equations(we set $\kappa=8\pi G=c=1$ in
the rest of this paper)
\begin{equation}\label{eq2}
G_{\mu \nu} = T^{({\rm eff})}_{\mu \nu},
\end{equation}
where the effective stress-energy tensor is
\begin{equation}\label{eq3}
T^{({\rm eff})}_{\mu \nu} = \frac{1}{\Phi}\left(T^{\rm m}_{\mu
\nu}+ T^{\Phi}_{\mu \nu}\right),
\end{equation}
with
\begin{equation}\label{eq4}
T^{\Phi}_{\mu\nu}=\frac{\omega}{\Phi}(\nabla_{\mu}\Phi\nabla_{\nu}\Phi-\frac{1}{2}g_{\mu\nu}\nabla^{\alpha}\Phi
\nabla_{\alpha}\Phi) +
(\nabla_{\mu}\nabla_{\nu}\Phi-g_{\mu\nu}\Box\Phi)
-\frac{1}{2}g_{\mu\nu}V(\Phi),
\end{equation}
and
\begin{equation}\label{eq5}
T^{m}_{\mu \nu} = {\rm
diag}\left(\rho_{m},p_{m},p_{m},p_{m}\right),
\end{equation}
being the stress-energy tensors of scalar field and matter fluid,
respectively. Here we look at the BD scalar field as matter fields
originating from geometry. Variation of the action with respect to
$\Phi$ gives
\begin{equation}\label{eq6}
\Box\Phi = \frac{T^{\rm m}}{2\omega + 3}+\frac{1}{2\omega +
3}\left(\Phi\frac{dV(\Phi)}{d\Phi}-2V(\Phi)\right).
\end{equation}
where $T^{m}$ stands for the trace of $T^{m}_{\mu \nu}$ and the
subscript "$m$" refers to the matter fields (fields other that
$\Phi$). In general relativity the interior solution of a
collapsing star is given by the Tolman-Bondi solution. Here we
present a transparent consideration of a homogenous collapsing
star with BD scalar field $\Phi = \Phi(\tau)$ and effective
potential $V=V(\Phi)$. With this consideration, the Tolman-Bondi
solution converts to a Friedmann-Robertson-Walker(FRW) metric. The
interior metric for marginally bound $(k=0)$ case is given by
\begin{equation}\label{eq7}
ds^2=-d\tau^{2} + a^{2}(\tau)(dr^{2} + r^{2}d\Omega^{2}),
\end{equation}
where $\tau$ is the proper time of free falling observer whose
geodesic trajectories are distinguished by the comoving radial
coordinate $r$ and $d\Omega^{2}$ being the standard line element
on the unit two sphere. Since the presence of matter acting as a
"seed" field and origin of spherical symmetry, prompts the
collapse of BD scalar field, we have considered perfect fluid
models with barotropic equation of state being given by
\begin{equation}\label{eq8}
p_{m} = w \rho_{m}.
\end{equation}
Using the continuity equation for the matter and Eq. (\ref{eq8}) one gets the
following relations between $\rho_{m} ,~ p_{m}$ and the scale factor as follows
\begin{equation}\label{eq9}
\rho_{m} =  \rho_{_{_{0}}m}a^{-3(1+w)};~~p_{m} =
w\rho_{_{_{0}}m}a^{-3(1+w)},
\end{equation}
where $\rho_{_{_{0}}m} = \rho_{m}(a=1)$, is the initial value of the energy density
of matter on the collapsing shell. One then has the following equations for the
effective stress-energy tensor
\begin{equation}\label{eq10}
\rho_{_{({\rm eff})}} =T^{\tau}\,_{\tau}^{({\rm eff})}=
\frac{1}{\Phi}\left(\rho_{\Phi} +
\rho_{m}\right)=\frac{1}{\Phi}\left(\rho_{m}+\frac{\omega}{2}\frac{\dot{\Phi}^2}{\Phi}-3\frac{\dot{a}}{a}\dot{\Phi}
+\frac{V(\Phi)}{2}\right),
\end{equation}
and
\begin{equation}\label{eq11}
p_{_{({\rm eff})}} =T^{r}\,_{r}^{({\rm
eff})}=T^{\theta}\,_{\theta}^{({\rm
eff})}=T^{\varphi}\,_{\varphi}^{({\rm eff})}=
\frac{1}{\Phi}\left(p_{_{\Phi}}
+p_{m}\right)=\frac{1}{\Phi}\left(p_{m}+\frac{\omega}{2}\frac{\dot{\Phi}^2}{\Phi}+\ddot{\Phi}+2\frac{\dot{a}}{a}\dot{\Phi}-\frac{V(\Phi)}{2}\right).
\end{equation}
with all other off-diagonal terms being zero and the radial and
tangential profiles of pressure are equal due to the homogeneity
and isotropy. The interior solution of Einstein's equation for the
line element (\ref {eq7}) takes the form
\begin{equation}\label{eq12}
\rho_{_{({\rm eff})}} = \frac{{\mathcal
M}^{\prime}}{R^{2}R^{\prime}};~~~~ p_{_{({\rm eff})}} =
-\frac{\dot{{\mathcal M}}}{R^{2}\dot{R}},
\end{equation}
\begin{equation}\label{eq13}
\dot{R}^{2}=\frac{{\mathcal M}}{R}.
\end{equation}
The quantity ${\mathcal M}$ arises as a free function from the
integration of Einstein's equation and can be interpreted
physically as the total mass of the collapsing cloud within a
coordinate radius r with ${\mathcal M}\geq0$,~and
$R(\tau,r)=ra(\tau)$ being the area radius for the shell labeled
by the comoving coordinate $r$. From Eq. (\ref{eq11}) we can solve
for the mass function as
\begin{equation}\label{eq14}
{\mathcal M}=\frac{R^3}{3\Phi}(\rho_{_{\Phi}} + \rho_{m}).
\end{equation}\
Using Eqs. (\ref{eq12}) and (\ref{eq13}) we arrive at a relation between $\dot{a}$
and the effective energy density as follows
\begin{equation}\label{eq15}
\dot{a}^2=\frac{a^2}{3\Phi}(\rho_{_{\Phi}} + \rho_{m}),
\end{equation}
since we are concerned with a continual collapse, the time variation of the scale
factor should be negative $(\dot{a}<0)$. This implies that the area radius of the
shell for constant value of $r$ decreases monotonically. For physical reasons, it
is assumed that the energy density is non-negative every where. The singularity
arising from continual collapse is given by $a=0$, in other words when the scale
factor and physical area radius of all the collapsing shells vanish, then the
collapsing cloud has reached a singularity. A point at which the energy density
blows up, the Kretschmann scalar $ {\mathcal K} = R^{abcd}R_{abcd}$ diverges
and the normal differentiability and manifold structures break down.
\section{The Solution}
We would like to construct and investigate a class of collapse
solutions for Brans-Dicke scalar filed with non-zero potential
considering the matter fields, where the trapping of light is
avoided till the singularity formation, thereby allowing the
singularity to be visible to outside observers. In order to reach
this purpose we consider a class of collapse models where near the
singularity, the divergence of energy density of BD scalar field
is given by the following \textit{ansatz}
\begin{equation}\label{eq16}
\rho_{_{\Phi}}=a^{-n},
\end{equation}
where $n$ is a positive constant and the scale factor, $a(\tau)$,
goes to zero in the limit of approach to the singularity which
causes $\rho_{_{\Phi}}$ to diverge. Using the above equation and
Eq. (\ref{eq15}), one may easily obtain the following relation for
$\ddot{a}$ as
\begin{equation}\label{eq17}
\ddot{a}=\frac{1}{6\Phi}\left[(2-n)a^{1-n}-(1+3w)\rho_{_{_{0}}m}a^{-(2+3w)}\right]-\frac{\Phi_{,a}}{6\Phi^2}\left[a^{2-n}+\rho_{_{_{0}}m}a^{-(1+3w)}\right],
\end{equation}
where $\Phi_{,a}=\dot{\Phi}/\dot{a}$. Now by substituting for
$\rho_{_{({\rm eff})}}$ and $p_{_{({\rm eff})}}$ into Eqs.
(\ref{eq10}) and (\ref{eq11}) together with the use of Eqs.
(\ref{eq13}), (\ref{eq15}) and (\ref{eq17}) we arrive at the
following differential equation as
\begin{align}\label{eq18}
&\frac{\Phi_{,a}}{\Phi^2}\left[ \left(\frac{n+2}{6}\right)a^{1-n}+
\rho_{_{_{0}m}}a^{-(2+3w)}\left(\frac{5+3w}{6}\right)\right]-\frac{\Phi_{,a}^2}{\Phi^3}\left[
\frac{2\omega+1}{6}\left(a^{2-n}+\rho_{_{_{0}m}}a^{-(1+3w)}\right)\right]\\
\nonumber
&+\frac{n}{3}\frac{a^{-n}}{\Phi}-\frac{\Phi_{,aa}}{3\Phi^2}\left[a^{2-n}+\rho_{_{_{0}m}}a^{-(1+3w)}\right]=0,
\end{align}
where we have used
\begin{equation}\label{eq19}
\dot{\Phi}=\dot{a}\Phi_{,a}~,~~~~~~~~~\ddot{\Phi}=\ddot{a}\Phi_{,a}+\dot{a}^2\Phi_{,aa}.
\end{equation}
In the following we solve Eq.~(\ref{eq18}) for barotropic equation
of state $p_{m} = w\rho_{m}$, where
$w=[0,-\frac{1}{3},-\frac{2}{3},\frac{1}{3}]$ correspond to dust,
cosmic string, domain wall, and radiation, respectively. Taking
the following ansatz
\begin{equation}\label{eq20}
\Phi(\tau) = a^{\alpha}(\tau),
\end{equation}
where $\alpha$ satisfies the following equation
\begin{align}\label{eq21}
-\alpha^2(1+\rho_{_{_{0}m}})(3+2\omega)+\alpha(n+4+\rho_{_{_{0}m}}(7+3w))+2n=0,
\end{align}
and by setting $n=3(1+w)$, we derive an expression for the BD
scalar field as a function of scale factor.
\subsection{Dust ($w=0$)}
For such a case of pressure-less matter the parameter $\alpha$
will take the following values as
\begin{align}\label{eq22}
\alpha=\left\{\begin{array}{c}
         -\frac{12}{7(1+\rho_{_{_{0}m}})+\sqrt{(1+\rho_{_{_{0}m}})(121+49\rho_{_{_{0}m}}+48\omega)}}
         \\\\
                \frac{12}{-7(1+\rho_{_{_{0}m}})+\sqrt{(1+\rho_{_{_{0}m}})(121+49\rho_{_{_{0}m}}+48\omega)}} \\
              \end{array}\right..
\end{align}

For $\rho_{_{_{0}m}}>0$ and $-\frac{3}{2}<\omega<\infty$ the first
value for $\alpha$ is negative and the second one is positive.
Thus, since the BD scalar field must diverge near the singularity
we choose the first one. For a special case in which
$\rho_{_{_{0}m}}=1$ and $\omega=-1$ (string effective action),
$\alpha=-0.405$.
\subsection{Cosmic Strings ($w=-\frac{1}{3}$)}
Cosmic strings are the consequence of 1-dimensional (spatially)
topological defects in various fields. These topological defects
are related to solitonic solutions of the classical equations for
the scalar (and gauge) fields which for the case of a complex
scalar field, cosmic strings can be formed\cite{30}. Such a kind
of matter can be regarded as a perfect fluid obeying the
barotropic equation of state, $p_{_{m}}=-1/3\rho_{_{m}}$. We are
interested here to find the behavior of BD scalar field as a
function of scale factor in gravitational collapse of this kind of
fluid. The corresponding values of $\alpha$ for this case are
\begin{align}\label{eq23}
\alpha=\left\{\begin{array}{c}
         -\frac{4}{3(1+\rho_{_{_{0}m}})+\sqrt{(1+\rho_{_{_{0}m}})(21+9\rho_{_{_{0}m}}+8\omega)}}
         \\\\
                \frac{4}{-3(1+\rho_{_{_{0}m}})+\sqrt{(1+\rho_{_{_{0}m}})(21+9\rho_{_{_{0}m}}+8\omega)}} \\
              \end{array}\right..
\end{align}
It is seen that for $\rho_{_{_{0}m}}>0$ and
$-\frac{3}{2}<\omega<\infty$ the first and second values of
$\alpha$ are negative and positive, respectively. Choosing the
first value and setting $\rho_{_{_{0}m}}=1$ and $\omega=-1$, we
have $\alpha=-0.316$.
\subsection{Domain Walls $(w=-\frac{2}{3})$}
Domain walls are Generally topological solitons which occur
whenever a discrete symmetry is spontaneously broken\cite{31}.
These are the result of 2-dimensional topological defects in
different scalar or gauge fields. As for the case of cosmic
strings, domain walls can be regarded as a perfect fluid obeying
the barotropic equation of state, $p_{_{m}}=-2/3\rho_{_{m}}$. We
here again try to find the behavior of BD scalar field as a
function of scale factor in gravitational collapse of such kind of
matter fluid. In order to reach this purpose we consider the
following values obtained for $\alpha$ as
\begin{align}\label{eq24}
\alpha=\left\{\begin{array}{c}
         \frac{4}{-5(1+\rho_{_{_{0}m}})+\sqrt{(1+\rho_{_{_{0}m}})(49+25\rho_{_{_{0}m}}+16\omega)}}
         \\\\
                -\frac{4}{5(1+\rho_{_{_{0}m}})+\sqrt{(1+\rho_{_{_{0}m}})(49+25\rho_{_{_{0}m}}+16\omega)}} \\
              \end{array}\right..
\end{align}
As it is seen the second one is always negative for
$\rho_{_{_{0}m}}>0$ and $-\frac{3}{2}<\omega<\infty$ causing the
BD scalar field to diverge in the vicinity of singularity. For
$\omega = -1$ and $\rho_{_{_{0}m}}=2$ one has $\alpha=-0.129$.
\subsection{Radiation $(w=\frac{1}{3})$}
Finally for this case of matter, we find the following values for
$\alpha$ as
\begin{align}\label{eq25}
\alpha=\left\{\begin{array}{c}
         -\frac{4}{2(1+\rho_{_{_{0}m}})+\sqrt{2(1+\rho_{_{_{0}m}})(5+2\rho_{_{_{0}m}}+2\omega)}}
         \\\\
                \frac{4}{-2(1+\rho_{_{_{0}m}})+\sqrt{2(1+\rho_{_{_{0}m}})(5+2\rho_{_{_{0}m}}+2\omega)}} \\
              \end{array}\right..
\end{align}
Again the first value is always negative for $\rho_{_{_{0}m}}>0$
and $-\frac{3}{2}<\omega<\infty$. Setting $\rho_{_{_{0}m}}=2$ and
$\omega = -1$ we have $\alpha=-0.32$.
\section{Time Behavior Of The Scale Factor}
One would like to study the time-dependence behavior of the scale
factor during the collapse procedure, considering matter. If at
time $\tau=\tau^{*}$ (or equivalently for some $a=a^{*}$) the
energy density of the BD scalar field starts growing as $a^{-n}$,
then by integrating Eq. (\ref{eq15}) in the vicinity of the
singularity with respect to time one gets the time behavior of the
scale factor as
\begin{equation}\label{eq26}
a(\tau)=\left(a^{\ast\frac{1}{2}\left(\alpha+3(1+w)\right)}-\frac{1}{2}\sqrt{\frac{1+\rho_{_{_{0}m}}}{3}}\left(\alpha+3(1+w)\right)(\tau-\tau_{\ast})\right)^{\frac{2}{\alpha+3(1+w)}},
\end{equation}
and the corresponding singular epoch as
\begin{equation}\label{eq27}
\tau_{s}=\frac{2\sqrt{3}}{\sqrt{1+\rho_{_{_{0}m}}}(\alpha+3(1+w))},
\end{equation}
where the time $\tau_{s}$ corresponds to a vanishing scale factor.
Thus the collapse reaches the singularity in a finite proper time.
This result for the scale factor completes the interior solution
within the collapsing cloud, providing us with the required
construction.
\section{Conditions On Radial Null Geodesic Expansion}
Consider a congruence of outgoing radial null geodesics having the
tangent vector $(\xi^{\tau},\xi^{r},0,0)$, where
$\xi^{\tau}=dt/d\lambda~and~\xi^{r} =dr/d\lambda$  and $\lambda$
is an affine parameter along the geodesics. In terms of these two
vector fields the geodesic equation can be written as\vspace{.7cm}
\begin{equation}\label{eq28}
\frac{d\xi^{r}}{d\lambda}=-\frac{2\dot{a}}{a}\xi^{t}\xi^{r},
\end{equation}
and\vspace{.7cm}
\begin{equation}\label{eq29}
\frac{d\xi^{t}}{d\lambda}=-a\dot{a}(\xi^{r})^{2}.
\end{equation}
The geodesic expansion parameter $\Theta$ is given by
\begin{equation}\label{eq30}
\Theta=\nabla_{j}\xi^{j}=\partial_{j}\xi^{j} +
\Gamma^{j}_{ji}\xi^{i},
\end{equation}
which gives
\begin{equation}\label{eq31}
\Theta=\frac{\partial\xi^{\tau}}{\partial \tau} +
\frac{\partial\xi^{r}}{\partial r}
+\left(\Gamma^{\tau}_{\tau\tau}+\Gamma^{r}_{r\tau}+\Gamma^{\theta}_{\theta
\tau}+\Gamma^{\Phi}_{\Phi \tau}\right)\xi^{\tau} +
\left(\Gamma^{\tau}_{\tau r}+\Gamma^{r}_{rr}+
\Gamma^{\theta}_{\theta r}+\Gamma^{\Phi}_{\Phi r}\right)\xi^{r}.
\end{equation}
In order to compute the sum \vspace{.7cm}
\begin{equation}\label{eq32}
\frac{\partial\xi^{\tau}}{\partial \tau} +
\frac{\partial\xi^{r}}{\partial r},
\end{equation}
we proceed by noting that\vspace{.7cm}
\begin{equation}\label{eq33}
\frac{d\xi^{\tau}}{d\lambda}=\frac{\partial\xi^{\tau}}{\partial
\tau} \frac{\partial \tau}{\partial\lambda} +
\frac{\partial\xi^{\tau}}{\partial r} \frac{\partial
r}{\partial\lambda},
\end{equation}
and similarly,\vspace{.7cm}
\begin{equation}\label{eq34}
\frac{d\xi^{r}}{d\lambda}=\frac{\partial\xi^{r}}{\partial \tau}
\frac{\partial \tau}{\partial\lambda} +
\frac{\partial\xi^{r}}{\partial r} \frac{\partial
r}{\partial\lambda}.
\end{equation}
Dividing the first of these two relations by $\partial \tau/\partial \lambda$ and the
second by $\partial r/\partial \lambda$, and after adding the resulted equations
one gets
\begin{equation}\label{eq35}
\frac{\partial\xi^{\tau}}{\partial \tau} +
\frac{\partial\xi^{r}}{\partial r}=
-\frac{1}{2}\left[\frac{2\dot{a}}{a}\xi^{\tau}+a\dot{a}\frac{(\xi^{r})^{2}}{\xi^{\tau}}\right]
+\frac{\dot{a}}{2a}\xi^{\tau},
\end{equation}
where we have used Eqs. (\ref{eq28}) and (\ref{eq29}), and the
fact that for outgoing radial null geodesics the relation between
$\xi^{t}$ and $\xi^{r}$ is given by\vspace{.7cm}
\begin{equation}\label{eq36}
\frac{\xi^{\tau}}{\xi^{r}}=\frac{d\tau}{dr}=a(\tau).
\end{equation}
Substituting Eq. (\ref{eq35}) into Eq. (\ref{eq31}) and after a
simple calculation we arrive at the desired expression for
$\Theta$:\vspace{.7cm}
\begin{equation}\label{eq37}
\Theta=\frac{2}{r}\left(1-\sqrt{\frac{{\mathcal
M}}{R}}~\right).\vspace{.4cm}
\end{equation}
In order to determine the visibility, or otherwise, of the
singularity, one needs to analyze the behavior of non-spacelike
curves in the vicinity of the singularity and the causal structure
of the trapped surfaces. These surfaces are closed orientable
smooth two-dimensional space-like surfaces such that both families
of ingoing and outgoing null geodesics orthogonal to them
necessarily converge \cite{32}. The singularity will be called
naked if there exists a family of future directed non-spacelike
geodesics, reaching faraway observers in space-time and
terminating at the singularity in the past. The existence of such
curves implies that either photons or time-like particles can be
emitted from singularity. So if the null geodesics terminate at
the singularity in the past with a definite tangent, then at the
singularity we have $\Theta>0 $. If such family of curves do not
exist and the event horizon forms earlier than the singularity
covering it, a blackhole is formed. The boundary of the trapped
surface region in the space-time is called apparent horizon where
in spherically symmetric space-time is given by
\begin{equation}\label{38}
g^{ik}R_{,i} R_{,k}=0.
\end{equation}
Therefore at the boundary of the trapped surface the vector
$R_{,k}$ is null. Using Eq. (\ref{eq7}), the above equation can be
written as
\begin{equation}\label{39}
-\dot{R}^{2} + a^{-2}R^{\prime 2}=0,
\end{equation}
which leads to ${\mathcal M} = R$. Here use has been made of Eq.
(\ref{eq13}). The space-time region where the mass function
${\mathcal M}$ satisfies ${\mathcal M}<R$ is not trapped, while
${\mathcal M}>R$ describes a trapped region \cite{33, 34, 04}.

Let us now study the relation that $\Theta$ bears with the
formation or otherwise of a naked singularity in spherical
collapse. Calculating ${\mathcal M}/R$ in the general case which
is considered for energy density $\rho_{_{\Phi}}$ and by using Eq.
(\ref{eq13}), we have
\begin{equation}\label{eq40}
\frac{{\mathcal M}}{R}=\frac{r^{2}}{3\Phi}\left(a^{2-n}
+\rho_{_{_{0}}m}a^{-(1+3w)}\right).
\end{equation}
We shall employ the above equation and Eq. (\ref{eq37}) to examine
the nakedness of the singularity as the collapse end state for the
four cases of matter field considered in section {\bf III}. We
show that physically, the formation of a black hole or a naked
singularity as the final state for the dynamical evolution is
governed by the rate of collapse scenario and the presence of BD
scalar field. It is seen that the cosmic censorship conjecture is
violated for $w=-\frac{1}{3}$ and $w=-\frac{2}{3}$. The weak
energy condition which states that the energy density as measured
by any local observer must be non-negative can be written for any
timelike vector $V^{\mu}$ as follows
\begin{equation}\label{eq41}
T_{\mu\nu}V^{\mu}V^{\nu}\geq0,
\end{equation}
whereby one gets the following conditions for the effective energy
density $(\rho_{_{({\rm eff})}}>0)$
\begin{equation}\label{eq42}
(1+\rho_{_{_{0}}m})a^{-(\alpha+3(1+w))}>0,
\end{equation}
and the sum of effective energy density and pressure
($\rho_{_{({\rm eff})}}+p_{_{({\rm eff})}}>0$) as
\begin{align}\label{eq43}
\left\{\begin{array}{c}
         w=0, ~\alpha<0 \rightarrow \left[(1+\rho_{_{_{0}}m})(1-\frac{\alpha}{3})\right]a^{-(\alpha+3)}>0, \\
         \\
         w=-\frac{1}{3}, ~\alpha<0 \rightarrow \left[(1+\rho_{_{_{0}}m})(\frac{4-\alpha}{3})\right]a^{-(\alpha+2)}>0, \\
         \\
         w=-\frac{2}{3}, ~\alpha<0 \rightarrow \left[(1+\rho_{_{_{0}}m})(\frac{5-\alpha}{3})\right]a^{-(\alpha+1)}>0, \\
         \\
         w=\frac{1}{3}, ~\alpha<0 \rightarrow \left[(1+\rho_{_{_{0}}m})(\frac{2-\alpha}{3})\right]a^{-(\alpha+4)}>0. \\
       \end{array}\right.
\end{align}
It is seen that these two conditions are satisfied by all cases of
$w$ and $\rho_{_{_{0}}m}>0$ and $-\frac{3}{2}<\omega<\infty$
considered above. At the initial epoch $(a=1)$ there should not be
any trapping of light. Assuming $r=r_{b}$ is the boundary of the
collapsing ball, then at the initial epoch the ratio ${\mathcal
M}/R$ is less than unity for a suitable initial value of
$\rho_{_{_{0}}m}$ standing for all values of $w$. This fact is in
accordance with the regularity condition stating that the
gravitational collapse must initiate from regular and physically
reasonable initial conditions. The time at which the physical area
radius of the collapsing cloud becomes zero, denotes a
shell-focusing singularity which lies on the curve
$R(\tau_{s},r)=0$ where $\tau_{s}$ being the singular epoch given
by Eq. (\ref{eq27}). For the case of homogeneous-density collapse
the resulting singularity may lay on the curves $R(\tau_{s},0)=0$
or $R(\tau_{s},r\neq0)=0$, which corresponds to a central or
non-central singularity, respectively. We consider first the
simpler case of non-central singularity and investigate the
failure of formation of apparent horizon in collapse scenario for
different values of $w$.
\subsection{Dust ($w=0$)}
We are now in a position to study the effect of BD scalar field on
the formation or otherwise of the apparent horizon as the
dynamical procedure of collapse scenario evolves( we set
$\omega=-1$ in the rest of this paper). We begin by Eq.
(\ref{eq14}) which for $w=0$ can be written as
\begin{equation}\label{eq44}
\frac{{\mathcal
M}}{R}=\frac{r^2}{3}(1+\rho_{_{_{0}}m})a^{-(1+\alpha)}.
\end{equation}
The initial energy density of matter must be positive due to the
regularity conditions, then for $\rho_{_{_{0}}m}>0$ and
$\omega=-1$, the first value of $\alpha$ in Eq. (\ref{eq22})
implies that $|\alpha|<1$. From Eq. (\ref{eq44}) it is seen that
the ratio ${\mathcal M}/R$ grows and the expansion parameter, Eq.
(\ref{eq37}), tends to negative infinity. Thus there exist no
radial null geodesics emerging from the singularity. Strictly
speaking the singularity occurred here is necessarily covered and
a black hole is formed as the collapse end state.
\subsection{Cosmic Strings ($w=-\frac{1}{3}$)}
For this case Eq. (\ref{eq14}) and the time variation of the mass
function take the following form as
\begin{align}\label{eq45}
\left\{\begin{array}{c}
         \frac{{\mathcal M}}{R}=\frac{r^2}{3}(1+\rho_{_{_{0}}m})a^{-\alpha}, \\
          \\
         \dot{{\mathcal M}}=-(1+\rho_{_{_{0}}m})\frac{r^3}{3}(\alpha-1)\dot{a}a^{-\alpha}. \\
       \end{array}\right.
\end{align}
As it is seen from Eq. (\ref{eq23}), since the first value of
$\alpha$ is always less than zero the ratio ${\mathcal M}/R$ stays
finite till the singular epoch and causes the expansion parameter
to be positive up to the singularity, and if no trapped surfaces
exist initially then no ones would form until the epoch
$a(\tau)=0$ which is consistent with the fact that there exist
families of outgoing radial null geodesics emerging from the
singularity. One can take the positive value of $\alpha$ in Eq.
(\ref{eq23}), but for this case the BD scalar field get vanished
as the scale factor tends to zero. Also the weak energy condition
may be violated. In addition to, for such a case the ratio
${\mathcal M}/R$ grows at a vanishing scale factor causing the
expansion parameter tends to negative infinity which means that
the singularity is covered and no radial geodesics can emerge from
it. From the second equation in Eq. (\ref{eq45}) it can be seen
that the time derivative of the mass function for $\alpha<0$, is
negative (note that $\dot{a}<0$) which means that the mass
function contained in the collapsing shell with that radius keeps
decreasing. In other words there exists an outward energy flux
during the collapse scenario. Since no trapped surfaces form up to
the singularity, the outward energy flux would be observable.
\subsection{Domain Walls ($w = -\frac{2}{3}$)}
For this case one may rewrite Eq. (\ref{eq14}) and time derivative
of the mass function as
\begin{align}\label{eq46}
\left\{\begin{array}{c}
         \frac{{\mathcal M}}{R}=\frac{r^2}{3}(1+\rho_{_{_{0}}m})a^{1-\alpha}, \\
          \\
         \dot{{\mathcal M}}=-(1+\rho_{_{_{0}}m})\frac{r^3}{3}(\alpha-2)\dot{a}a^{1-\alpha}. \\
       \end{array}\right.
\end{align}
from the first equation one may easily see that at initial epoch
($a = 1$), the regularity condition is satisfied. Since the second
value of $\alpha$ in Eq. (\ref{eq24}) is always negative, the
ratio of mass function to area radius of the collapsing shell is
less than unity during the collapse procedure denoting that the
expansion parameter being positive up to the singularity. In this
case the collapse evolution to a naked singularity takes place,
where the trapped surfaces do not form early enough or are avoided
in the spacetime. For first value of Eq. (\ref{eq24}), $\alpha>5$
which causes the ratio ${\mathcal M}/R$ tends to infinity as the
scale factor vanishes and $\Theta$ goes to negative infinity, thus
trapped surfaces do form in the spacetime which prevent the null
geodesics to emerge from the singularity. Such a situation ends in
a black hole as the final fate of the collapse scenario. But such
a value of $\alpha$ is not allowed since it violates the weak
energy condition. From the second equation in Eq. (\ref{eq46}), it
is obvious that for negative value of $\alpha$ and $\dot{a}<0$,
the time derivative of the mass function is negative stating that
the mass contained in collapsing ball reduces  as the time
advances.
\subsection{Radiation ($w=\frac{1}{3}$)}
In this case Eq. (\ref{eq14}) can be written as
\begin{equation}\label{eq47}
\frac{{\mathcal
M}}{R}=\frac{r^2}{3}(1+\rho_{_{_{0}}m})a^{-(\alpha+4)}.
\end{equation}
From Eq. (\ref{eq25}) one can easily see that the first value for
$\rho_{_{_{0}}m}$ and $\omega=-1$ is always negative and
$|\alpha|<4$. Thus, in such a situation the ratio ${\mathcal M}/R$
tends to infinity as the singularity is approached. Thus the
expansion parameter behaves just as the dust case, and the final
singularity is necessarily covered within an event horizon of
gravity.

The central singularity occurring at $R=0,r=0$ is naked if there exist outgoing
non-spacelike geodesics reaching faraway observers and terminating in the past
at the singularity. In order to investigate the nakedness of this kind of singularity
we proceed by introducing a new variable $x=r^{\delta},~ and ~ \delta>1$ is
defined such that $R^{\prime}/r^{\delta-1}$ is a unique finite quantity in the limit
$r\rightarrow0$. Then we have the following equation
\begin{equation}\label{eq48}
\frac{dR}{dx}=\frac{1}{\alpha
r^{\alpha-1}}\left(\dot{R}\frac{d\tau}{dr}+R^{\prime}\right).
\end{equation}
By virtue of Eqs. (\ref{eq13}) and (\ref{eq36}) the above equation
leads to
\begin{equation}\label{eq49}
\frac{dR}{dx}=\frac{R^{\prime}}{\alpha
x^{\frac{\alpha-1}{\alpha}}}\left[1-\sqrt{\frac{{\mathcal
M}}{R}}~\right].
\end{equation}
It is clear that $R=0$ , $x=0$ is a singular point of Eq.
(\ref{eq49}). If there are outgoing radial null geodesics
terminating in the past at the singularity with a definite
tangent, then at the singularity we have $\frac{dR}{dx}>0$. For
$w=-\frac{1}{3}$, and $w=-\frac{2}{3}$  with $\alpha$ being
negative the quantity ${\mathcal M}/R<1$ throughout the collapse
procedure, so the term being in the second bracket is positive and
$\frac{dR}{dx}>0$ as the singularity is approached indicating that
the singularity is visible to outside observers and the inverse
result holds for $w=0$ and $w=\frac{1}{3}$.
\section{nakedness of the singularity}
The continual gravitational collapse of a matter cloud culminates
in either a black hole or a naked singularity where in the former
an event horizon develops earlier than the formation of the
singularity. Thus the regions of extreme physical conditions such
as densities and curvatures are hidden from the outside observers.
If such horizons are delayed or failed to develop during the
collapse procedure, as governed by the internal dynamics of the
collapsing cloud, then the scenario where the ultra-strong gravity
regions become visible to external observers occurs and a visible
naked singularity forms. In such a case where no black hole forms,
the field collapses for a while and then disperses. Therefore as
viewed by a central observer, the scalar invariants namely the
Kretschmann scalar should grow near the singularity, gain some
maximum value and then approach to zero at late times\cite{35}.
Since the absence of an apparent horizon does not necessarily
implies the absence of an event horizon, we examine the nakedness
of the singularity in spherically symmetric collapse of a fluid by
considering the behavior of the Kretschmann invariant with respect
to time. For the line element (7) this quantity is given by
\begin{equation}\label{eq50}
{\cal K}\equiv
R^{abcd}R_{abcd}=\frac{12}{a^4}\left[a^2\ddot{a}^2+\dot{a}^4\right].
\end{equation}
By the virtue of Eqs. (\ref{eq26}) and (\ref{eq27}) for
$w=-\frac{1}{3}$ the above quantity can be written as
\begin{equation}\label{eq51}
{\cal K}_{cs}=\frac{3.322}{(1-0.6\tau)^{4}},
\end{equation}
and for $w=-\frac{2}{3}$ as
\begin{equation}\label{eq52}
{\cal K}_{dw}=\frac{11.6}{(1-0.4\tau)^{4}}.
\end{equation}
Note that both these two cases satisfy the condition on expansion
parameter stating that this quantity must be positive up to the
singularity. Fig. 1 and Fig. 2 show the behavior of the
Kretschmann scalar as a function of proper time, $\tau$. It is
seen that both ${\cal K}_{cs}$ and ${\cal K}_{dw}$ diverge at
$\tau=5/3$ and $\tau=5/2$, respectively. They then converge to
zero at late times signaling the failure of formation of the event
horizon.

Let us now consider the geometry of the exterior spacetime. In
order to fully complete the spacetime model, one needs to match
the interior spacetime of the dynamical collapse to a suitable
exterior geometry. The Schwarzschild solution is a useful model
describing the spacetime outside the sun and stars. However this
model may no longer be suitable to describe the exterior geometry
of any realistic star, because the spacetime outside such a star
may be filled with radiated energy from the star in the form of
electromagnetic radiation. The Schwarzschild model does not
describe this as it corresponds to an empty spacetime given by
$T_{ab}=0$. The spacetime outside a spherically symmetric star
being surrounded by a radiation emitted from the star is described
by the Vaidya metric\cite{36} which can be given in the form
\begin{equation}\label{eq53}
ds^{2}_{out}=-\left(1-\frac{2M(r_{u},u)}{r_{u}}\right)du^{2} -
2dudr_{u} +r_{u}^2d\Omega^2,
\end{equation}
where \textit{u}, being the retarded null coordinate, $r_{u}$ and
$M(r_{u},u)$ are the Vaidya radius and Vaidya mass, respectively.
Following the work of \cite{33} we use the Isreal-Darmois junction
conditions to match the interior spacetime described by Eq.
(\ref{eq7}) to a Vaidya exterior geometry at the boundary
hypersurface $\Sigma$ given by $r = r_{b}$. The spacetime metric
just inside $\Sigma$ is given by
\begin{equation}\label{eq54}
ds^{2}_{in}=-d\tau^2+a^2(\tau)\left[dr^2+r_{b}^2d\Omega^2\right]
\end{equation}
Matching the area radius of the collapsing shell at the boundary,
one gets the following equation
\begin{equation}\label{eq55}
r_{u}(u)=r_{b}a(\tau),
\end{equation}
whereby on the hypersurface $\Sigma$, the interior and exterior
metrics can be written as
\begin{equation}\label{eq56}
ds^{2}_{\Sigma in}=-d\tau^2+a^2(\tau)r_{b}^2d\Omega^2,
\end{equation}
and
\begin{equation}\label{eq57}
ds^{2}_{\Sigma out}=-\left(1-\frac{2M(r_{u},u)}{r_{u}} +
2\frac{dr_{u}}{du}\right)du^{2}+r_{u}^2d\Omega^2.
\end{equation}
Matching the induced metric on $\Sigma$ one gets
\begin{equation}\label{eq58}
\left(\frac{du}{d\tau}\right)_{\Sigma}=\frac{1}{\left(1-\frac{2M(r_{u},u)}{r_{u}}+2\frac{dr_{u}}{du}\right)^\frac{1}{2}},~~~
(r_{u})_{\Sigma}=r_{b}a(\tau).
\end{equation}
In order to match the extrinsic curvature for interior and
exterior spacetimes, one has to fine the unit normal vector field
to the hypersurface $\Sigma$. In this step, we proceed by noting
that any spacetime metric can be written locally in the form
\begin{equation}\label{eq59}
ds^2=-\left(N^2-N_{i}N^{i}\right)d\tau^2-2N_{i}dx^{i}d\tau+h_{ij}dx^{i}dx^{j},
\end{equation}
where N, $N^{i}$, and $h_{ij}$ are the lapse function, shift
vector, and induced metric, respectively and $i$, $j$ are
three-dimensional indices run in $\{1,2,3\}$. The contravariant
and covariant components of the unit normal vector field are given
by
\begin{equation}\label{eq60}
n^{a}=\frac{1}{N}\left(\delta^{a}_{0}-N^{a}\right),~~~n_{a}=-N\delta^{0}_{a}.
\end{equation}
Comparing Eqs. (\ref{eq59}) and (\ref{eq7}), one finds the
contravariant components of the normal to the hypersurface
$\Sigma$ for the interior metric as
\begin{equation}\label{eq61}
n^{a}_{in}=[0,a(\tau)^{-1},0,0].
\end{equation}
Upon using a similar approach, one finds the first non-vanishing
contravariant component of the normal to $\Sigma$ for the exterior
metric as
\begin{equation}\label{eq62}
n^{u}=-\frac{1}{\left(1-\frac{2M(r_{u},u)}{r_{u}}+2\frac{dr_{u}}{du}\right)^\frac{1}{2}}.
\end{equation}
In order to compute the second non-vanishing contravariant
component of the normal vector field we proceed by having recourse
the normalization relation holding for $n^{a}$ as
\begin{equation}\label{eq63}
n^{u}n_{u} + n^{r_{u}}n_{r_{u}}=1.
\end{equation}
Benefiting from the property of the metric tensor in raising and
lowering indices, we have the following relations
\begin{equation}\label{eq64}
n_{u}=\frac{1-\frac{2M(r_{u},u)}{r_{u}}}{\left(1-\frac{2M(r_{u},u)}{r_{u}}+2\frac{dr_{u}}{du}\right)^\frac{1}{2}}-n^{r_{u}};~~~~
n_{r_{u}}=-n^{u}.
\end{equation}
Substituting the above equations and Eq. (\ref{eq62}) into Eq.
(\ref{eq63}) and after a simple calculation we arrive at the
desired expression for $n^{r_{u}}$ as
\begin{equation}\label{eq65}
n^{r_{u}}=\frac{1-\frac{2M(r_{u},u)}{r_{u}}+\frac{dr_{u}}{du}}{\left(1-\frac{2M(r_{u},u)}{r_{u}}+2\frac{dr_{u}}{du}\right)^\frac{1}{2}}.
\end{equation}
The extrinsic curvature of the hypersurface $\Sigma$ is defined as
the Lie derivative of the metric tensor with respect to the normal
vector field being given by the following relation
\begin{equation}\label{eq66}
K_{ab}=\frac{1}{2}\left[g_{ab,c}n^{c}+g_{cb}n^{c}_{,a}+g_{ac}n^{c}_{,b}\right].
\end{equation}
Since the matching is for the second fundamental form, $K_{ab}$,
there exists no surface stress energy or surface tension at the
boundary\cite{37}. The nonzero $\theta$ components of the
extrinsic curvature are
\begin{equation}\label{eq67}
K^{in}_{\theta\theta}=r_{b}a(\tau),~~~~~~K^{out}_{\theta\theta}=r_{u}\frac{1-\frac{2M(r_{u},u)}{r_{u}}+\frac{dr_{u}}{du}}{\left(1-\frac{2M(r_{u},u)}{r_{u}}+2\frac{dr_{u}}{du}\right)^\frac{1}{2}}.
\end{equation}
Setting
$\left[K^{in}_{\theta\theta}-K^{out}_{\theta\theta}\right]_{\Sigma}=0$
on the hypersurface $\Sigma$, and by using Eqs. (\ref{eq13}) and
(\ref{eq58}), one gets the following relation between mass
function and Vaidya mass on the boundary as
\begin{equation}\label{eq68}
{\mathcal M}(\tau,r_{b})=2M(r_{u},u).
\end{equation}
From the above equation and Eq. (\ref{eq14}) one can see that the
BD scalar field affects on the behavior of the Vaidya mass in the
collapse scenario. In order to find another relation expressing
the rate of change of Vaidya mass with respect to $r_{u}$ one has
to match the $\tau$ component of the extrinsic curvature on the
hypersurface $\Sigma$. Having set
$\left[K^{in}_{\tau\tau}-K^{out}_{\tau\tau}\right]=0$, one gets
\begin{equation}\label{eq69}
M(r_{u},u)_{r_{u}}=\frac{{\mathcal M}}{2r_{u}}+r_{b}^2a\ddot{a}.
\end{equation}
The occurrence of a naked singularity as the final outcome of a
collapse procedure, depends on the non-existence of trapped
surfaces till the formation of the singularity, which corresponds
to the existence of families of non-spacelike trajectories
reaching faraway observers and terminating in the past at the
singularity. In order to show this, we begin by Eq. (\ref{eq58})
and after using Eqs. (\ref{eq67}), and (\ref{eq68}) we arrive at
the following relation
\begin{equation}\label{eq70}
\left(\frac{du}{d\tau}\right)_{\Sigma}=\frac{1-r_{b}\dot{a}}{1-\frac{{\mathcal
M}(\tau,r_{b})}{r_{u}}}.
\end{equation}
It can be easily checked that if one imposes the null condition on
the Vaidya metric, the result is the same as Eq.~\ref{eq70}. What
is meant by this, is that null geodesics can come out from the
singularity and reach to faraway observers before it evaporates
into the free space. In other words the formation of trapped
surfaces in spacetime is avoided and a naked singularity can be
produced.
 \section{Behavior Of The BD Scalar field and the Effective Potential}
In the following section we wish to study how the effective
potential behaves as the scalar field varies. For this purpose we
start by Eq. (\ref{eq10}) and consider the four cases of matter
field discussed in section {\bf III}. Together with the use of
Eqs. (\ref{eq15}) and (\ref{eq20}), one may easily find the
following expression for the potential as
\begin{align}\label{eq71}
V(\phi)=\beta\Phi^{-\frac{3(1+w)}{\alpha}},~~~~~\beta=\left(2+\frac{\alpha}{3}(1+\rho_{_{_{0}}m})(6-\omega)\right).
\end{align}
Fig. 3 shows the behavior of the BD potential with respect to
$\Phi$ for different values of $w$ and $\omega=-1$.

Let us now consider the case in which the BD scalar field is a
function of both $\tau$ and $r$. Assuming that far away from the
collapsing system, the effective energy density behaves
homogeneously, we obtain a measure of the radial profile of the BD
scalar field for each cases of $w$ considered in Section {\bf
III}. We begin by Eq. (\ref{eq6}) together with the use of Eqs.
(\ref{eq15}), (\ref{eq17}), (\ref{eq19}), and (\ref{eq71}) we
arrive at a differential equation for $\Phi(a(\tau),r)$ as
\begin{align}\label{eq72}
\left[\frac{n-8}{6}a^{1-n}+\frac{\rho_{_{_{0}m}}}{6}(3w-5)a^{-(3w+2)}\right]\frac{\phi_{,a}}{\phi}+
&\frac{1}{6}\left[a^{2-n}+\rho_{_{_{0}m}}a^{-(1+3w)}\right]\left(\frac{\phi_{,a}}{\phi}\right)^2-\left[a^{2-n}+\rho_{_{_{0}m}}a^{-(1+3w)}\right]\frac{\phi_{,aa}}{3\phi}\\\nonumber
&+\frac{\phi^{\prime\prime}}{a^2}+2\frac{\phi^{\prime}}{ra^2}+C\frac{a^{-3(1+w)}}{2\omega+3}=0,
\end{align}
where $C$ is given by
\begin{align}\label{eq73}
C=\frac{3\beta(1+w)}{\alpha}+2\beta-\rho_{_{_{0}m}}(3w-1).
\end{align}
the above equation can be more simplified and the result is as
follows
\begin{align}\label{eq74}
\phi^{\prime\prime}+\frac{2\phi^{\prime}}{r}-Da^{1-\alpha-3w}=0,
\end{align}
where $D$ is a constant and is given by
\begin{align}\label{eq74}
D=\frac{\alpha(\alpha-1)}{3}+\left(\frac{5\alpha-\alpha^2}{6}-\frac{\alpha
w}{2}\right)(1+\rho_{_{_{0}m}}) -\frac{C}{2\omega+3}.
\end{align}
Here $\prime$ denotes partial differentiation with respect to $r$.
Solving the above differential equation with a suitable choice of
value for $\rho_{_{_{0}}m}$ one may find the solutions as
functions of $\tau$ and $r$. Figs. 4-7 show the behavior of BD
scalar field in terms of $\tau$ and $r$ in which the constants of
integration have been chosen in such a way that the scalar field
blows up near the singularity and get vanished faraway  from the
collapsing system.
\section{Conclusion and outlook}
In this work we have studied the gravitational collapse of the BD
scalar field with non-zero potential in the presence of matter
fluid. Assuming that the energy density of the BD scalar field
behaves as the inverse power law of the scale factor near the
singularity, we presented a class of solutions in Brans-Dicke
theory in which the naked singularity can be created being
accompanied by the violation of the cosmic censorship conjecture.
In section \textbf{III}, we found the behavior of BD scalar field
as a function of scale factor for the four cases of matter fluid.
In section \textbf{IV}, the general expressions of time behavior
of the scale factor and singular epoch has been achieved. Having
examined the \textit{ansatz} taken for divergence of the energy
density of BD scalar field, Eq. (\ref{eq16}), together with the
use of Eq. (\ref{eq9}) for energy density of matter fluid and by
using the concept of expansion parameter, we have shown in section
\textbf{V} that the presence of BD scalar field due to the Eq.
(\ref{eq14}) can affect the formation or otherwise of the trapped
surfaces and only for the two cases, $w=-\frac{1}{3}$ and
$w=-\frac{2}{3}$, formation of the apparent horizon can be failed
and a naked singularity may be generated as the final fate of
collapse procedure. But since the absence of an apparent horizon
does not necessarily implies the absence of an event horizon, we
have computed the Kretschmann scalar in section \textbf{VI} and
the result has been plotted in Fig. 1 and Fig. 2 for
$w=-\frac{1}{3}$ and $w=-\frac{2}{3}$, respectively. It is seen
that this quantity diverges at singular epoch, and then vanishes
at late times, a behavior which can be interpreted as the absence
of an event horizon and formation of a naked singularity. Also
following the work of \cite{33} we have shown at the end of this
section that the Vaidya geodesic emerging from the singularity
before it evaporates into free space is null.

Beside our work which have only treated exact solutions to
gravitational collapse in Brans-Dicke theory, one may find
numerical solutions to such a collapse scenario in the
literature\cite{38}. In \cite{25, 29}, the author has developed a
new numerical code that solves the gravitational field equations
coupled to the matter for evolution of a spherically symmetric
configuration of noninteracting particles in Brans-Dicke theory.
Using this code, he has shown that Oppenheimer-Snyder collapse in
this theory results in black holes rather than naked
singularities, at least for $|3+2\omega|\geq3$, which are
identical to those of general relativity in final equilibrium, but
are quite different from those of general relativity during
dynamical evolution in which they radiate mass. The reason for
this behavior is due to the violation of the null energy condition
even in vacuum spacetimes with positive values of $\omega$,
passing of the apparent horizon of a black hole outside the event
horizon, and decreasing the surface area of the event horizon over
time. This numerical code enables one to decide on a number of
long-standing theoretical questions about collapse in Brans-Dicke
theory of gravitation. Also the gravitational collapse of a scalar
field with other characteristics and couplings has been discussed
in some literature. In \cite{39}, the collapse of a self-similar
scalar field has been studied, and it has shown that there exists
two classes of solutions which one of them consists of a
nonsingular origin in which the scalar field collapses and
disperses again. There is a singularity at one point of these
solutions which is not observable at a finite radius. The second
class of solutions contains both black holes and naked
singularities with a critical behavior interpolating between these
two extremes. Numerical study of spherically symmetric collapse of
a massless scalar field has presented in \cite{40}, where it is
shown that the masses of black holes which form satisfy a power
law $M_{BH}\propto|p-p^{\ast}|^{\gamma}$. Where $p$ is a parameter
which characterizes the strength of initial condition, $p^{\ast}$
is the threshold value and $\gamma\approx0.37$ is a universal
exponent. Also the collapse of a massless scalar field in
Brans-Dicke theory has studied both analytically and numerically
in \cite{41} and it is shown that for $\omega>-\frac{3}{2}$, a
continuous self-similarity continues and that the critical
exponent $\gamma$ depends on $\omega$. In \cite{42}, gravitational
collapse of a self-interacting (massive) scalar field has been
studied both analytically and numerically on a
Reissner-Nordsr$\ddot{o}$m background and finally in \cite{43},
one may find some examples of naked singularity formation in
gravitational collapse of a scalar field.

\section*{Acknowledgments}
Y. Tavakoli is supported by the Portuguese Agency Funda\c{c}\~{a}o
para a Ci\^{e}ncia e Tecnologia through the fellowship
SFRH/BD/43709/2008.

\begin{figure}
\begin{center}
\epsfig{figure=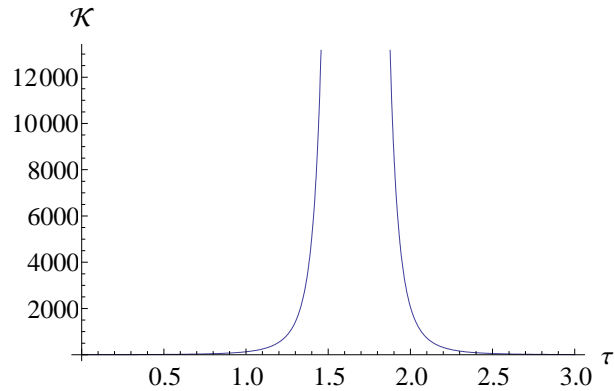,width=8cm}
\end{center}
\caption{\footnotesize The behavior of Kretschmann scalar (in
units of $s^{-4}$) as a function of proper time for $\omega=-1$
and $w=-\frac{1}{3}$. For the initial energy density, scale
factor, and proper time we have adopted the values
$\rho_{_{0m}}=1$, $a^{\ast}=1$, and $\tau^{\ast}=0$,
respectively.}
\end{figure}
\begin{figure}
\begin{center}
\epsfig{figure=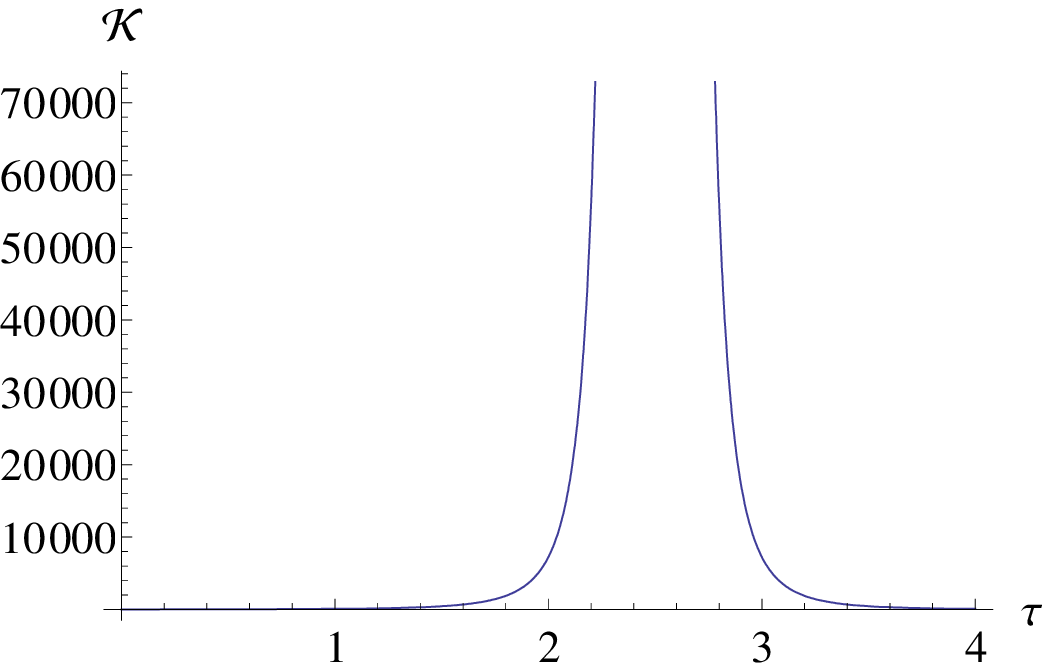,width=8cm}
\end{center}
\caption{\footnotesize The behavior of Kretschmann scalar (in
units of $s^{-4}$) as a function of proper time for $\omega=-1$
and $w=-\frac{2}{3}$. For the initial energy density, scale
factor, and proper time we have adopted the values
$\rho_{_{0m}}=2$, $a^{\ast}=1$, and $\tau^{\ast}=0$,
respectively.}
\end{figure}
\begin{figure}
\begin{center}
\epsfig{figure=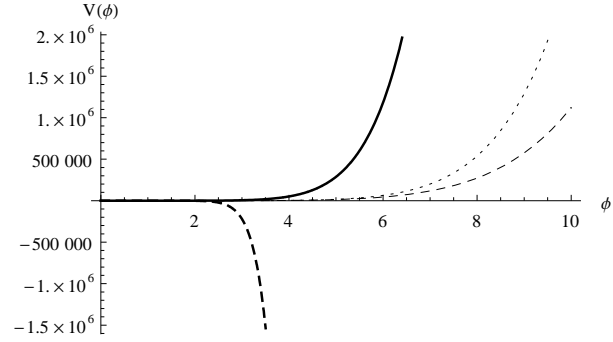,width=8cm}
\end{center}
\caption{\footnotesize  Behavior of potential as a function of BD
scalar field for $\omega=-1$ and different values of
$\rho_{_{_{0}m}}$ and $w$: $\rho_{_{_{0}m}}=1$ and $w=0$ (Dotted
curve), $\rho_{_{_{0}m}}=1$ and $w=-\frac{1}{3}$ (Dashed Curve),
$\rho_{_{_{0}m}}=2$ and $w=-\frac{2}{3}$ (Solid Curve),
$\rho_{_{_{0}m}}=2$ and $w=\frac{1}{3}$ (Thick-Dashed Curve).}
\end{figure}
\begin{figure}
\begin{center}
\epsfig{figure=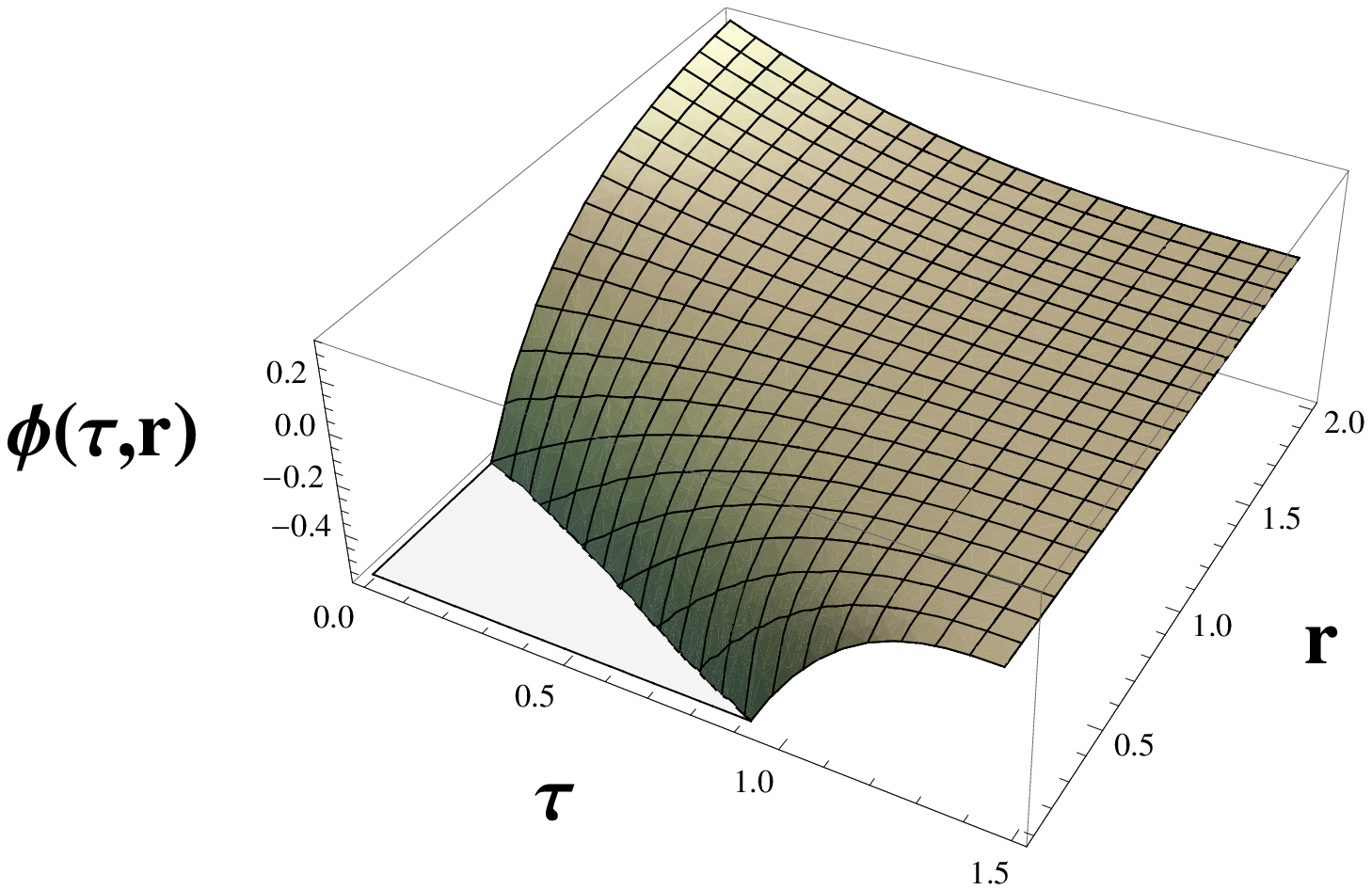,width=8cm}
\end{center}
\caption{\footnotesize Behavior of the BD scalar field with
respect to $\tau$ and r for $\omega =-1$, $w=-\frac{1}{3}$. For
the initial energy density, scale factor and proper time we have
adopted the values $\rho_{_{_{0}m}} = 1$, $a* = 1$ and $t* = 0$,
respectively.}
\end{figure}
\begin{figure}
\begin{center}
\epsfig{figure=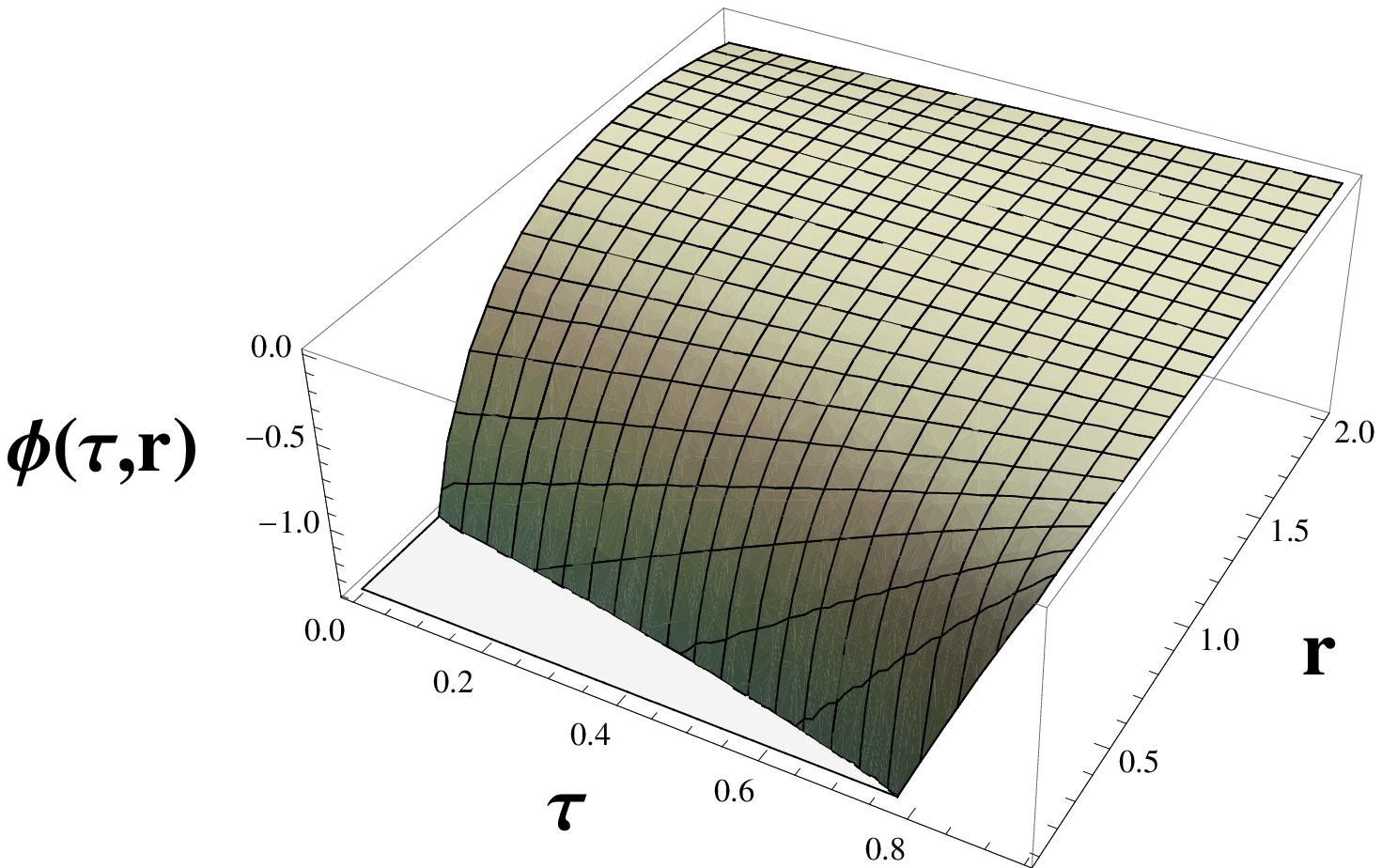,width=8cm}
\end{center}
\caption{\footnotesize Behavior of the BD scalar field with
respect to $\tau$ and r for $\omega =-1$, $w=0$. For the initial
energy density, scale factor and proper time we have adopted the
values $\rho_{_{_{0}m}} = 1$, $a* = 1$ and $t* = 0$,
respectively.}
\end{figure}
\begin{figure}
\begin{center}
\epsfig{figure=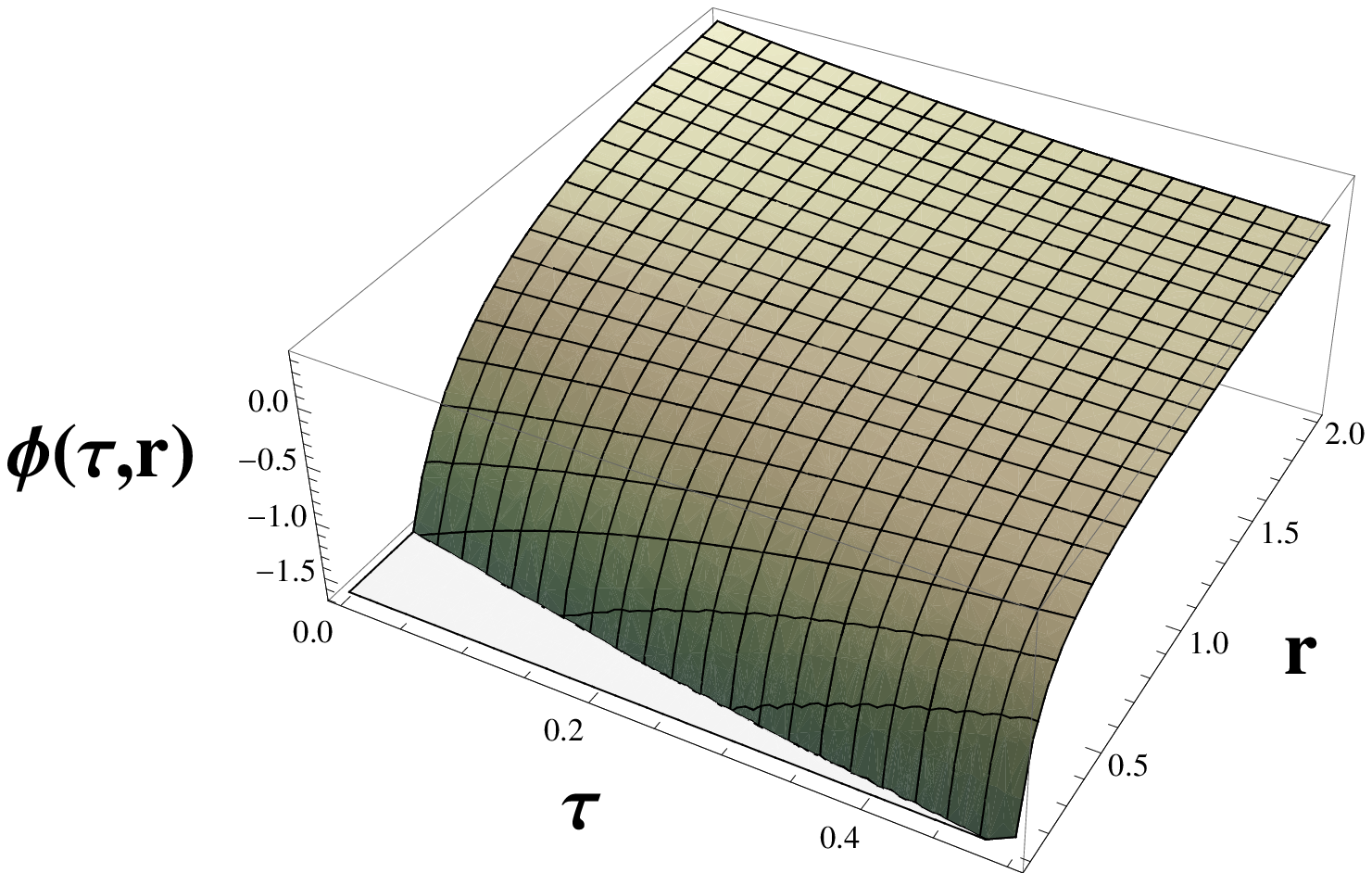,width=8cm}
\end{center}
\caption{\footnotesize Behavior of the BD scalar field with
respect to $\tau$ and r for $\omega =-1$, $w=-\frac{2}{3}$. For
the initial energy density, scale factor and proper time we have
adopted the values $\rho_{_{_{0}m}} = 2$, $a* = 1$ and $t* = 0$,
respectively.}
\end{figure}
\begin{figure}
\begin{center}
\epsfig{figure=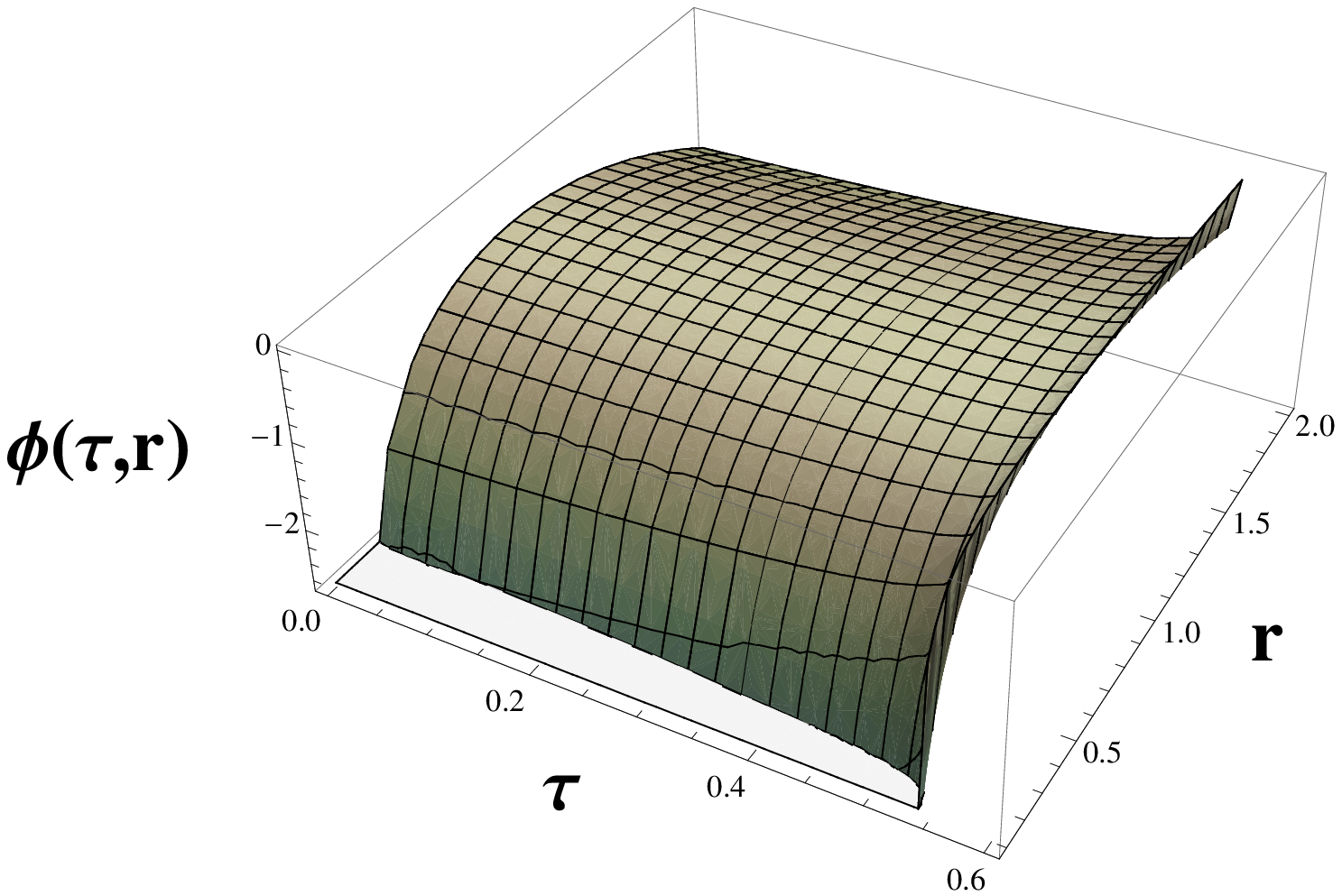,width=8cm}
\end{center}
\caption{\footnotesize Behavior of the BD scalar field with
respect to $\tau$ and r for $\omega =-1$, $w=\frac{1}{3}$. For the
initial energy density, scale factor and proper time we have
adopted the values $\rho_{_{_{0}m}} = 2$, $a* = 1$ and $t* = 0$,
respectively.}
\end{figure}

\begin{thebibliography}{99}

\bibitem{01} P. S. Joshi, {\it Gravitational Collapse and Space-Time Singularities},
(Cambridge University Press, 2007).
\bibitem{02} A. Gromov, Yu. Baryshev, D. Suson and P. Teerikorpi, gr-qc/9906041.
\bibitem{03} G. ${\it Lema\hat{i}tre}$ Ann. Soc. Sci Bruxelles A \textbf{53} (1933) 51.
\bibitem{04} P. S. Joshi and I. H. Dwivedi, Phys. Rev. D \textbf{47} (1993) 5357-5369, gr-qc/9303037.
\bibitem{05} R. C. Tolman, Proc. Natl. Acad. Sci. USA \textbf{20} (1934) 169.
\bibitem{06} H. Bondi, Mon. Not. Astron. Soc. \textbf{107}  (1947) 410.
\bibitem{07} A. Ori and T. Piran, Phys. Rev. D \textbf{42} (1990) 1068.
\bibitem{08} P. Yodzis, H. Seifert and H. $M\ddot{u}ller$ zum Hagen, Commun. Math. Phys. \textbf{34}  (1973) 135.
\bibitem{09} D. M. Eardly and L. Smarr, Phys. Rev. D \textbf{19} (1979) 2239 ;\\
D. Christodoulou, Commun. Math. Phys. \textbf{93} (1984) 171.
\bibitem{10} W. A. Hiscock, L. G. Williams and D. M. Eardley, Phys. Rev. D \textbf{26}
(1982) 751.
\bibitem{11} R. Goswami and P. S. Joshi, Phys. Rev. D \textbf{69} (2004) 027502, gr-qc/ 0310122.
\bibitem{12} J. R. Oppenheimer and H. Snyder, Phys. Rev. \textbf{56} (1939) 455.
\bibitem{13} D. Christodoulou , gr-qc/0805.3880.
\bibitem{14} R. Penrose, Rev. Nuovo Cimento \textbf{1} (1969) 252.
\bibitem{15} F. C. Mena, B. C. Nolan and R. Tavakol, Phys. Rev. D \textbf{70} (2004) 084030, gr-qc/0405041.
\bibitem{16} D. Christodoulou, Ann. of Math. \textbf{149} (1999) 183;
\\J. M. Martin-Garcia  and C. Gundlach , Phys. Rev. D \textbf{68} (2003) 024011.
\bibitem{17} K. D. Patil, Phys. Rev. D \textbf{67} (2003) 024017.
\bibitem{18} P. S. Joshi and T. P. Singh, Phys. Rev. D \textbf{51} (1995) 6778;
\\ I. H. Dwivedi and P. S. Joshi, Class. Quant. Grav. \textbf{14} (1997) 1223;
\\ S. H. Ghate, R. V. Saraykar and K. D. Patil, Pramana, J. Phys. \textbf{53} (1999) 253.
\bibitem{19} Y. Kuroda, Prog. Theor. Phys. \textbf{72} (1984) 63; \\K. Rajagopal and K. Lake, Phys. Rev. D \textbf{35} (1987) 1531;\\
I.
H. Dwivedi and P. S. Joshi, Class. Quant. Grav. \textbf{6} (1989) 1599;\\
P. S. Joshi and I. H. Dwivedi, Gen. Relativ. Gravit. \textbf{24} (1992) 129;\\ J.
Lemos, Phys. Rev. Lett. \textbf{68} (1992) 1447;
\\K. D. Patil,
R.V. Saraykar and S. H. Ghate, Pramana, J. Phys. \textbf{52} (1999) 553;\\
S. G. Ghosh and R. V. Saraykar, Phys. Rev. D \textbf{62} (2000) 107502.
\bibitem{20} P. S. Joshi and I. H. Dwivedi, Commun. Math. Phys. \textbf{146}
(1992) 333;\\ T. Harada, Phys. Rev. D \textbf{58} (1998) 104015.
\bibitem{21} P. Szekeres and V. Lyer, Phys. Rev. D \textbf{47}
(1993) 4362;\\ S. Brave, T. P. Singh and L. Witten, Gen. Relativ. Gravit.
\textbf{32} (2000) 697.
\bibitem{22} S. G. Ghosh and N. Dadhich, Gen. Relativ. Gravit. \textbf{35}
(2003) 359;\\ T. Harko  and S. K. Cheng, Phys. Lett. A \textbf{266} (2000) 249.
\bibitem{23} P. S. Joshi, R. Goswami and N. Dadhich, gr-qc/0308012.
\bibitem{24} P. S. Joshi, N. Dadhich and R. Maartens, Phys. Rev. D
\textbf{65} (2000) 101501, gr-qc/0109051.
\bibitem{25} M. A. Scheel, S. L. Shapiro, and S. A. Teukolsky, gr-qc/9411026.
\bibitem{26} P. G. O Freund, Nucl. Phys. B \textbf{209} (1982) 146.
\bibitem{27} M. B. Green, J. H. Schwarz, and E. Witten,
Superstring Theory: 2 (Cambridge: Cambridge Univercity Press,
1987); C. G. Callan, D. Friedan, E. J. Martinek, and M. J. Perry,
Nucl. Phys. B \textbf{262} (1985) 593.
\bibitem{28} D. La and P. J. Steinhardt, Phys. Rev. Lett.
\textbf{62} (1989) 376; P. J. Steinhardt and F. S. Accetta, Phys. Rev. Lett.
\textbf{64} (1990) 2740; J. Garcia-Bellido and M. Quiros, Phys. Lett. B
\textbf{243} (1990) 45.
\bibitem{29} M. A. Scheel, S. L. Shapiro, and S. A. Teukolsky, gr-qc/9411025.
\bibitem{30} V. Mukhanov, {\it Phisical Foundations of Cosmology},
(Cambridge University Press, 2005).
\bibitem{31} S. Weinberg, {\it The Quantum Theory of Fields}, Vol. 2. Chap 23, Cambridge University Press (1995).
\bibitem{32} V. P. Frolov and I. D. Novikov, {\it Black Hole Physics} , Copenhagen
$\&$ Edmonton, October 1997.
\bibitem{33} R. Goswami and P. S. Joshi, Mod. Phys. Lett. A \textbf{22} (2007) 65.
\bibitem{34} P. S. Joshi and R. Goswami, Phys. Rev. D \textbf{69} (2004)
064027, gr-qc/0206042.
\bibitem{35} D. Garfinkle, and G. Comer Duncan, Phys. Rev. D
\textbf{58} (1998) 064024, gr-qc/9802061.
\bibitem{36} P. C. Vaidya, {\it The External Field of a Radiating Star In General
Relativity}, Curr. Sci. \textbf{12} (1943) 183; P. C. Vaidya, {\it
The Gravitational Field of a Radiating Star}, Proc. Indian Acad.
Sci. A. \textbf{33} (1951) 264; P. C. Vaidya, {\it Newtonian Time
In General Relativity}, Nature, \textbf{171} (1953) 260.
\bibitem{37} P. O. Mazur, and E. Mottola, (2004).'Gravitational vacuum condensate stars'.
Proc. Nat. Acad. Sci., \textbf{111}, 9545.
\bibitem{38} K. S. Thorne and J. J. Dykla, Ap. J. \textbf{166},
L35 (1971); S. W. Hawking, Comm. Math. Phys. \textbf{25} (1972)
167; T. Matsuda, Prog. Theo. Phys. \textbf{47} (1972) 738.
\bibitem{39} P. R. Brady, Phys. Rev. D \textbf{51} (1995) 4168.
\bibitem{40} M. W. Choptuik, Phys. Rev. Lett. \textbf{70} (1992) 9.
\bibitem{41}T. Chiba, J. Soda, Prog. Theor. Phys. \textbf{96} (1996)
567, gr-qc/9603056.
\bibitem{42} S. Hod, T. Piran, Phys, Rev. D \textbf{58} (1998)
044018, gr-qc/9801059.
\bibitem{43} D. Christodoulou, Annals, Math. \textbf{140} (1994)
607.
\end{thebibliography}
\end{document}